\documentclass[lettersize,journal]{IEEEtran}
\usepackage{amsmath,amsfonts}
\usepackage{algorithmic}
\usepackage{array}
\usepackage[caption=false,font=normalsize,labelfont=sf,textfont=sf]{subfig}
\usepackage{textcomp}
\usepackage{stfloats}
\usepackage{url}
\usepackage{verbatim}
\usepackage{graphicx}
\usepackage{xcolor}
\def\BibTeX{{\rm B\kern-.05em{\sc i\kern-.025em b}\kern-.08em
    T\kern-.1667em\lower.7ex\hbox{E}\kern-.125emX}}
\usepackage{balance}
\begin{document}
\title{\fontsize{22pt}{26pt}\selectfont VIKIN: A Reconfigurable Accelerator for KANs and\\ MLPs with Two-Stage Sparsity Support}
\author{Wenhui Ou, Zhuoyu Wu, Yipu Zhang,  Zheng Wang and C. Patrick Yue
\thanks{
This work was supported by Hong Kong Research Grants Council through the Areas of Excellence (AoE) Scheme under Grant AoE/E-601/22-R and NSFC under Grant No. 62372442. (\textit{Corresponding author: C. Patrick Yue}.)

Wenhui Ou, Yipu Zhang, and C. Patrick Yue are with the Department of Electronic and Computer Engineering, The Hong Kong University of Science and Technology, Hong Kong SAR, China.

Zhuoyu Wu is with the School of IT, Monash University, Malaysia Campus, and Zheng Wang is with the Shenzhen Institutes of Advanced Technology, Chinese Academy of Sciences, Shenzhen, China.
}}

\maketitle

\begin{abstract}
\noindent The commonly used Multi-layer Perceptrons (MLPs) in modern AI applications significantly limit the real-time performance due to their intensive memory access operation. Recently, Kolmogorov–Arnold Networks (KANs) have gained attention for offering a similar structure to MLPs but with a more efficient parameter utilization. However, the lack of customized support in conventional hardware restricts the performance gain from its algorithmic superiority. Furthermore, the limited applicability of KANs in edge scenarios, especially where MLPs excel, makes dedicated hardware for KANs inefficient and impractical. In this work, we present VIKIN, a reconfigurable accelerator that enables efficient inference for both KANs and MLPs using unified hardware, providing flexibility to handle diverse edge workloads. Apart from its pipeline mode and two-stage sparsity support for efficient KAN processing, VIKIN can also improve the throughput of MLPs with the same sparsity support in parallel mode. Experiments on a real-world dataset show that VIKIN provides a 1.28$\times$ acceleration and a 19.58\% reduction in accuracy loss by replacing MLPs with KANs. For a more accurate KAN model version with 3.29$\times$ operation, the latency overhead on VIKIN is only 1.24$\times$ compared to the baseline KAN model. Additionally, VIKIN achieves a 1.25$\times$ speed-up and 4.87$\times$ energy efficiency compared to a common edge GPU when performing KAN model.
\end{abstract}

\vspace{-2mm}

\section{Introduction}
\IEEEPARstart{M}{u}lti-layer Perceptrons (MLPs) are widely adopted and serve as fundamental building blocks in advanced neural networks, due to their simple structure and strong versatility. However, the same factors that make MLPs attractive also lead to inherent limitations. First, the poor weight reuse and the large number of parameters required for satisfactory accuracy impose significant pressure on memory bandwidth, making MLPs a bottleneck on hardware. For example, in a specialized Transformer accelerator \cite{tu2022trancim}, MLPs account for up to 60\% of the total latency and energy consumption. Second, MLPs are essentially black-box models with limited interpretability~\cite{radenovic2022neural}, making it difficult to obtain feedback for network optimization after training. These limitations motivate the exploration of emerging alternatives.

\begin{figure}[h]
  \centering
  \includegraphics[width=1\columnwidth]{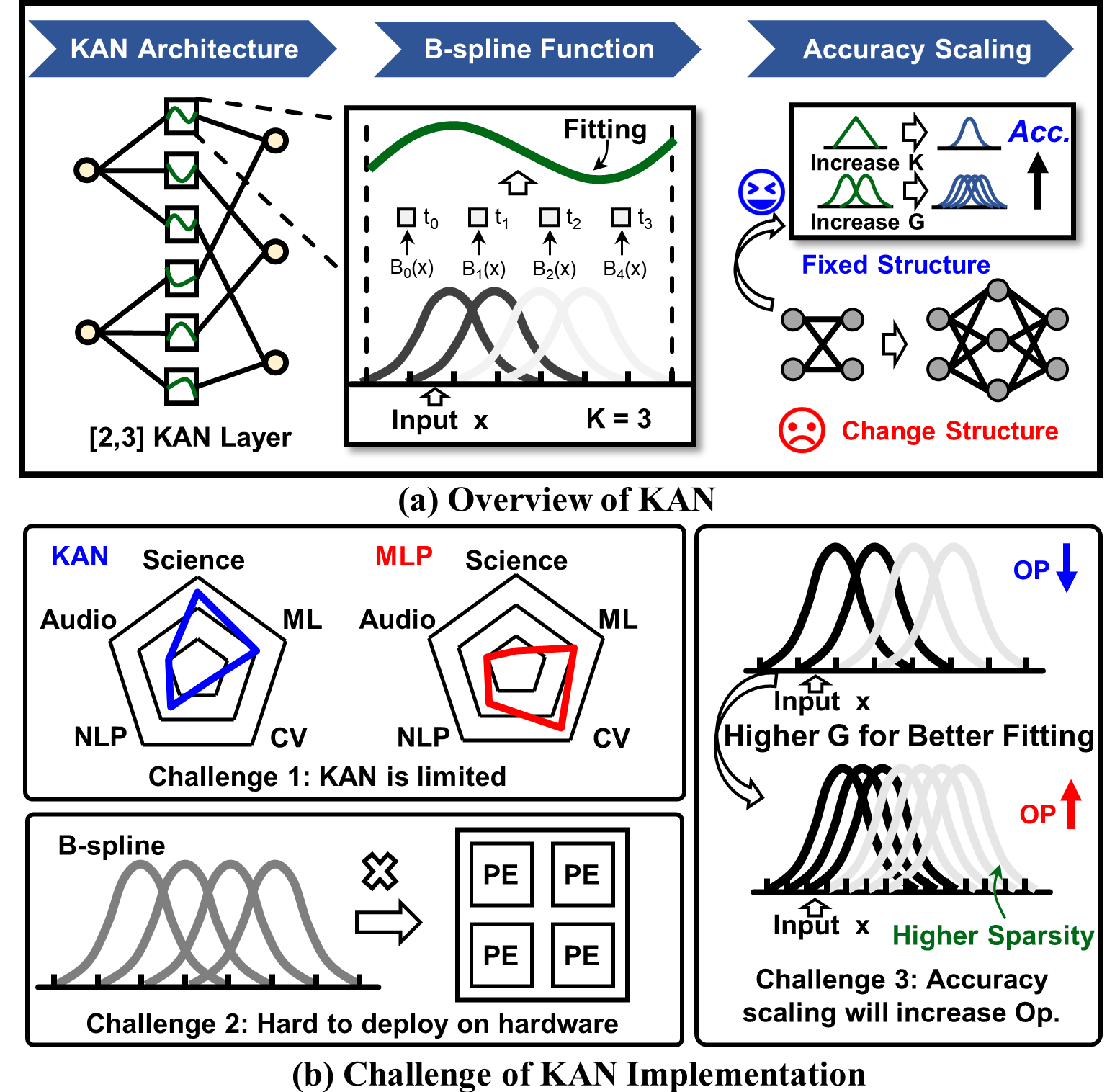}
    \caption{(a) Overview of KAN and (b) its implementation challenges. In (a), we show an example of a [2,3] KAN layer, where the output is constructed by linearly combining multiple B-spline basis functions with learnable weights. The use of B-splines enables accuracy scaling by tuning spline parameters instead of changing the model structure. In (b), the comparison between KAN and MLP is summarized based on~\cite{yu2024kan}, highlighting the need for unified hardware that can efficiently support both models.}
  \label{Fig1:KAN}
  \vspace{-5mm}
\end{figure}

Recently, Kolmogorov–Arnold Networks (KANs) \cite{liu2024kan} have emerged as a competitive alternative to MLPs. Specifically, KANs retain the same high-level architecture as MLPs, making them a drop-in replacement in different neural networks (NNs). Moreover, in certain cases, KANs can deliver higher accuracy while using fewer parameters compared to MLPs \cite{liu2024kan}. Most importantly, by introducing B-spline functions \cite{gordon1974b} as key components, KANs not only improve interpretability but also enable a smoother accuracy-scaling mechanism, where accuracy can be enhanced by tuning internal parameters instead of redesigning the overall NN architecture, as shown in Fig.~\ref{Fig1:KAN}(a). Recent studies have verified that the integration of KANs can improve the performance of mainstream models such as Transformers~\cite{yang2024kolmogorov} and LSTMs~\cite{karnehm2024core}, further demonstrating their potential for broader deployment.

Nevertheless, Bringing the algorithmic benefits of KANs to hardware, especially edge devices in diverse workload environments, still faces several challenges, as shown in Fig.~\ref{Fig1:KAN}(b): (1) According to the latest survey \cite{yu2024kan}, the benefits of KANs are confined to specific applications, such as scientific tasks, while their performance in other domains remains limited, indicating that KANs cannot yet fully replace MLPs across all workloads; (2) Existing high-parallelism hardware accelerators are typically based on processing element (PE) arrays~\cite{dong202528nm}, which are well-optimized for MLPs but struggle to support KANs due to the unique computations involved in B-spline functions; (3) Although KANs allow smoother accuracy scaling without changing the architecture, they also introduce significant operation overhead, which can sometimes exceed those of MLPs and may counteract the efficiency gains.

To address the aforementioned challenges, we make the following contributions in this paper:

\begin{itemize}
    \item We propose \textbf{VIKIN}, a versatile inference engine that can operate in two modes: a pipeline mode for KANs, and a parallel mode for MLPs, to adapt to varying workload characteristics with unified hardware.
    
    \item We design a reconfigurable B-spline Unit (SPU) that can efficiently execute B-spline functions with different parameter settings, while it can also be reused for MLPs to enhance hardware utilization.
    
    \item We integrate dedicated sparsity support to optimize KAN's accuracy-scaling capability, which can significantly reduce computational overhead of more accurate KAN models, while MLPs can also benefit from this.
    
    \item Experimental results show that, compared with a common edge GPU, KAN achieves a $1.25\times$ speedup and a $4.87\times$ improvement in energy efficiency on our dedicated engine, whereas MLPs also benefit from our design, achieving a $2.20\times$ energy saving.
\end{itemize}

\section{Preliminary}
\subsection{Kolmogorov Arnold Networks}
\label{subsection: KAN}
\noindent KAN was derived from the Kolmogorov-Arnold representation theorem \cite{hecht1987kolmogorov}, which was originally composed of two layers. A KAN layer $\phi_{q,p}$ with $n_{in}$ inputs and $n_{out}$ outputs \cite{liu2024kan} can be defined as:

\begin{equation}
\begin{aligned}
\Phi= \{ \phi_{q,p} \},  \:p=1,2,..,n_{in}, \: q=1,2,...,n_{out}
\end{aligned}
\label{eq:kanlayer}
\end{equation}

The KAN layers can be generalised in Eq.~\ref{eq:kanequation1}, $\text{where }\text{silu}(x) = x/(1 + e^{-x}) \text{ and spline}(x)=\sum_{i=0}^{m} c_{i}B_{i}(x)$, with learnable weights $w_b$ and $w_s$. Using the similar fully-connected structure of MLPs, KANs can be expanded to any $L$ layers with the output expressed as KAN($x$) = $\Phi _{L-1}\circ\Phi _{L-2}\cdot \cdot\cdot\circ\Phi _{0}\circ x_0$ \cite{liu2024kan}.

\begin{equation}
\begin{aligned}
\phi(x)= w_{b}\text{silu}(x)+w_{s}\text{spline}(x)
\end{aligned}
\label{eq:kanequation1}
\end{equation}

For hardware-friendly implementation, the learnable weight $w_s$ can be combined with $c_i$ to form $t_i = w_sc_i$

\begin{equation}
\begin{aligned}
\phi(x)= w_{b}\text{silu}(x)+\overset{m}{\underset{i=0}{\varSigma} }t_{i}B_{i}(x)
\end{aligned}
\label{eq:kanequation2}
\end{equation}

Fig.~\ref{Fig1:KAN}(a) illustrates a KAN layer with 2 inputs and 3 outputs (denoted as $[2,3]$). The computation can be conceptually divided into two stages: 
(1) the SiLU activations and B-spline basis functions $B_i(x)$ are first computed; 
(2) the resulting intermediate values are processed via multiply-and-accumulate (MAC) operations using learnable weights $w_b$ and $t_i$. Since each input generates $(m{+}1)$ intermediate result in the first stage, the second stage can be viewed as a linear layer of an MLP with dimensions $[2 \cdot (m{+}1), 3]$. Here, $m$ represents the number of B-spline basis, as shown in Eq.~\ref{eq:kanequation2}.

\subsection{B-Spline Functions and Accuracy Scaling}

\noindent KAN employs B-spline functions to approximate nonlinear mappings by combining localized basis components $B_i(x)$ with trainable coefficients $t_i$ (Fig.~\ref{Fig1:KAN}(a)) \cite{gordon1974b}. A B-spline basis is specified by the spline order $K$ and the knot/grid positions $\{x_i\}$ (grid size $G$ determines the number of knot placements within a fixed input interval). The zero-order basis is
\begin{equation}
\begin{aligned}
B_{0,i}(x)=
\begin{cases}
1 & x_i \le x < x_{i+1},\\
0 & \text{otherwise}
\end{cases}
\end{aligned}
\label{eq:zero-order basis}
\end{equation}

and higher-order bases can be recursively computed from lower-order ones:
\begin{equation}
\begin{split}
B_{K,i}(x)=&
\frac{x-x_{i}}{x_{i+K}-x_{i}}\,B_{K-1,i}(x)\\ 
&\quad + \frac{x_{i+K+1}-x}{x_{i+K+1}-x_{i+1}}\,B_{K-1,i+1}(x)
\end{split}
\label{eq:higher-order basis}
\end{equation}

This parametrization allows accuracy to be scaled in two ways, as shown in Fig.~\ref{Fig1:KAN}(a): using a larger $K$ for smoother and more expressive bases, or increasing $G$ to obtain more fine-grained local fitting, all without changing the model architecture. In particular, enlarging only $G$ offers a lightweight way to boost accuracy without retraining the model from scratch~\cite{liu2024kan}. Although a larger $G$ increases the number of $B_i(x)$ evaluations, it also leads to higher sparsity, since $B_i(x)$ becomes zero when $x$ lies outside its local support, as illustrated in Fig.~\ref{Fig1:KAN}(b). This sparsity can be effectively exploited to reduce computational overhead.

\section{Overall Architecture}
\begin{figure}[h]
  \centering
  \includegraphics[width=0.9\columnwidth]{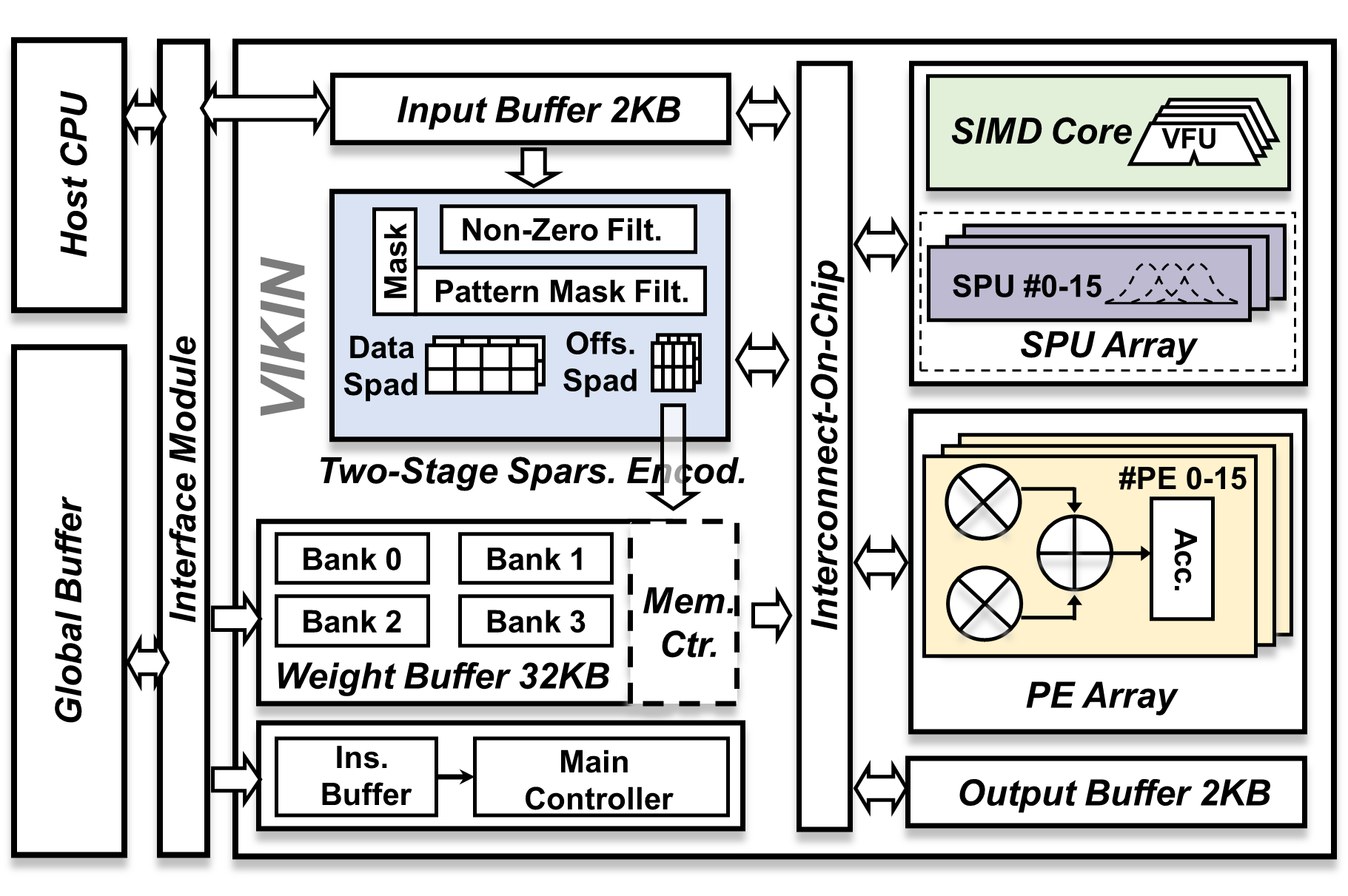}
  \caption{Overall hardware architecture of VIKIN.}
  \label{Fig2:overall_architecture}
\end{figure}

\noindent \noindent Fig.~\ref{Fig2:overall_architecture} shows the overall architecture of VIKIN. Like most accelerators \cite{hong2022efficient}, it operates alongside a host processor that handles configuration and issues flexible control instructions. A global buffer stores most weight data and provides high bandwidth, avoiding frequent off-chip DDR access and ensuring sufficient data supply for inference.  

\begin{figure}[h]
  \centering
  \includegraphics[width=1\columnwidth]{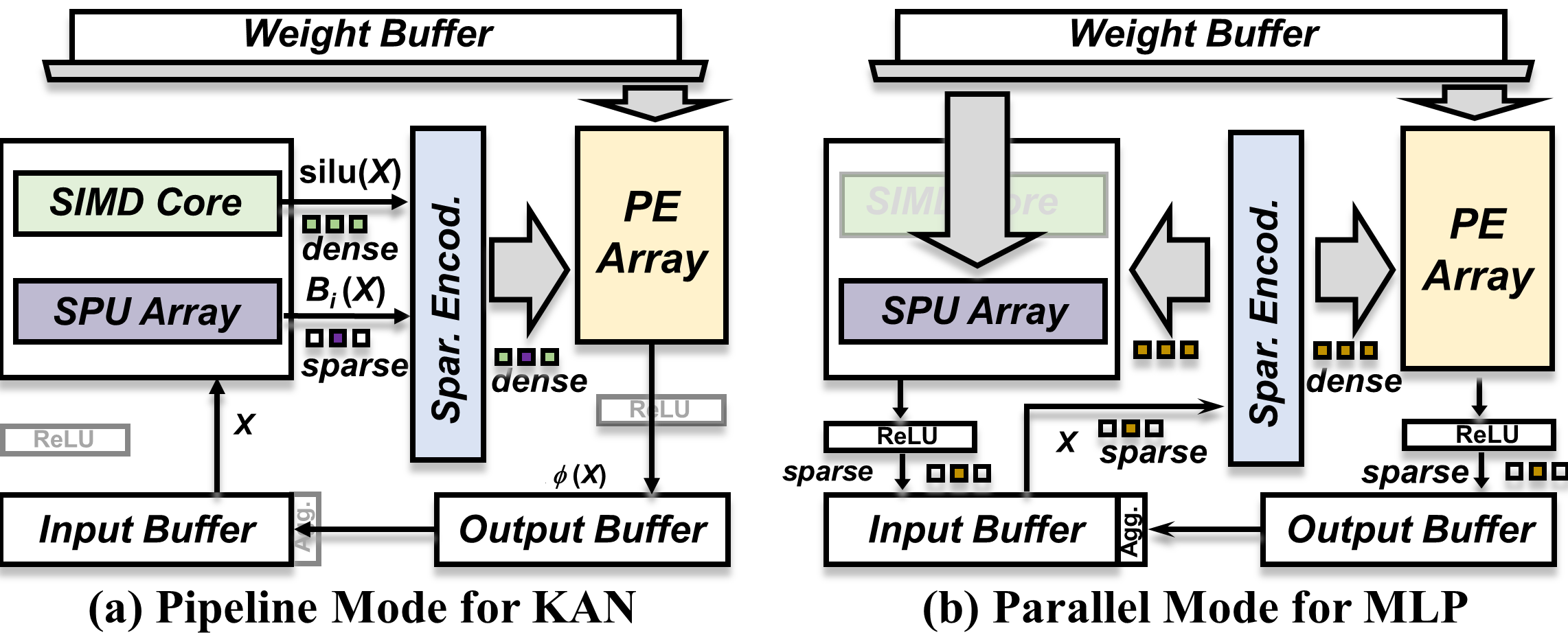}
  \caption{The illustration of the reconfigurable dataflow: (a) pipeline mode for KAN, (b) parallel mode for MLP.}
  \label{Fig3:reconfigurable_dataflow}
  \vspace{-4mm}
\end{figure}

The core of VIKIN includes a SIMD core~\cite{ou2023compact} for silu($x$), a 16-unit B-spline unit (SPU) array for computing $B_i(x)$, and a 16-unit processing element (PE) array for MAC operations, as outlined in Eq.~\ref{eq:kanequation2}. Additionally, a two-stage sparsity encoder (TSE) supports sparsity-aware computation in both parallel and pipeline modes. To optimize performance for accuracy-sensitive applications~\cite{koenig2024kan,cheon2024kolmogorov}, KAN inference is performed in FP16, striking a balance between speed and accuracy.

\begin{figure}[h]
  \centering
  \includegraphics[width=1\columnwidth]{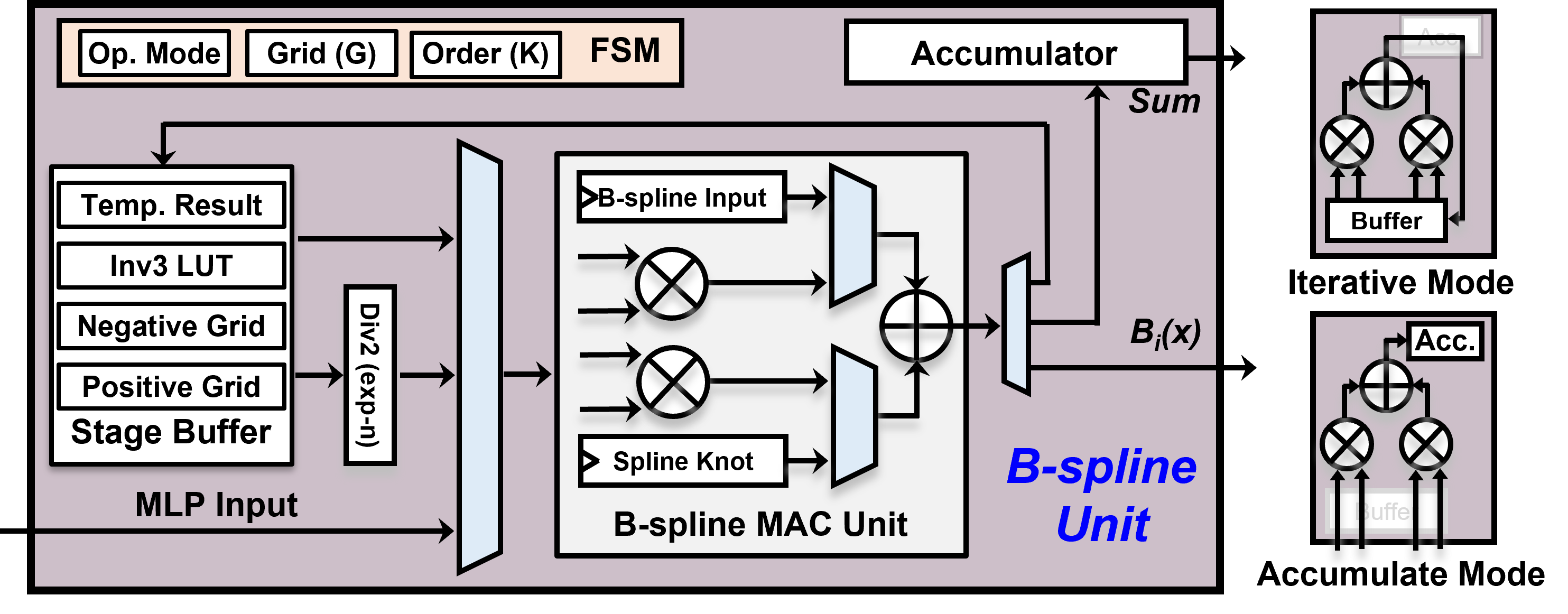}
  \caption{The hardware architecture of the reconfigurable B-spline unit (SPU) supports two modes: iterative mode for B-spline bases $B_i(x)$ in KAN, and accumulation mode, which functions as a PE for MLP.}
  \label{Fig4:spline_unit}
\end{figure}

\begin{figure*}[t]
  \centering
  \includegraphics[width=0.9\textwidth]{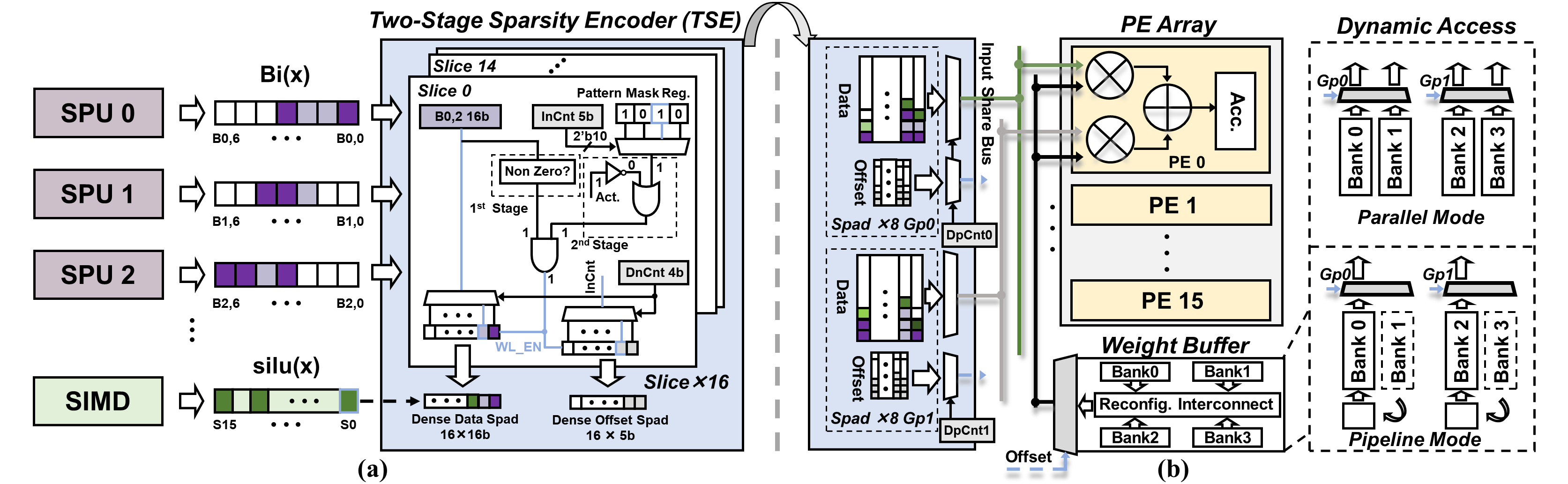}
    \caption{The pipeline mode with sparsity support: (a) pipeline stage 1, (b) pipeline stage 2. This mode hides SPU latency through pipeline processing and reduces PE array operations via sparsity-aware computation in the TSE. The figure also illustrates the dynamic weight buffer access scheme across different operation modes to ensure sufficient bandwidth.}

  \label{Fig5:pipeline mode}
    \vspace{-3mm}
\end{figure*}

\vspace{-3mm}
\section{Reconfigurable Mode and Microarchitecture}

\subsection{Reconfigurable Dataflow}
\label{subsection:reconfigurable dataflow}
\noindent VIKIN supports two operation modes. In pipeline mode for KANs, as shown in Fig.~\ref{Fig3:reconfigurable_dataflow}(a), the SIMD core and SPU array process input data from the input buffer. Since $B_i(x)$ is bounded, the SPU array produces sparse outputs with many zeros. The TSE forwards only the non-zero $B_i(x)$ results to the PE array, which then writes outputs directly to the output buffer, bypassing the ReLU block.

In parallel mode for MLPs (Fig.~\ref{Fig3:reconfigurable_dataflow}(b)), the interconnect reconfigures the hardware to connect the input buffer directly to the TSE. The SPU array will function like PE array to perform MAC operations. SPU outputs are stored back in the input buffer, while PE outputs go to the output buffer before aggregation into the input buffer for the next layer. Reusing the TSE converts sparse MLP inputs from preceding ReLU activations into a dense format, enabling efficient sparsity-aware computation with minimal reconfiguration overhead.

\vspace{-2mm}
\subsection{Reconfigurable B-spline Unit}
\label{subsection:reconfigurable B-spline Unit}
\noindent The hardware implementation of B-spline functions faces two main challenges. First, computing B-spline bases incurs high operation overhead (Eq.~\ref{eq:higher-order basis}). Although LUTs can speed up computation, their performance is limited by the need to refill them whenever layer configurations change ($G$ and $K$), resulting in high memory access costs, especially in dynamic workloads. Second, the unique computation paradigm of B-splines results in highly customized hardware, which can lead to low utilization if it cannot be reused for other operations.

To improve efficiency, the SPU design incorporates several optimizations, as shown in Fig.~\ref{Fig4:spline_unit}: (1) We follow the computation paradigm of EfficientKAN~\cite{efficientKAN} with higher speed. (2) Differences between the input and knots (i.e., $x-x_i$ and $x_{i+K+1}-x$ in Eq.~\ref{eq:higher-order basis}) are computed once for the zero-order (K=0) basis and stored in a stage buffer, then reused for higher-order bases, thereby reducing workload by 21\%. (3) Additionally, $G$ is restricted to $2, 4, 8, 16$ and $K$ to $1, 2, 3, 4$, allowing costly FP divisions in Eq.~\ref{eq:higher-order basis} to be replaced with integer operations and an LUT for the value $1/3$.

To enable reuse for MLPs and improve hardware utilization, SPUs support two modes. In iterative mode, SPUs compute higher-order $B_i(x)$ by fetching lower-order ones from the stage buffer. In accumulation mode for MLP operations, SPUs receive external inputs and use accumulators to perform MAC. Compared to the 16 nodes per batch in KAN pipeline mode, using the SPU array doubles the number of output nodes per batch for MLPs, allowing up to 32 nodes.

\vspace{-2mm}
\subsection{Pipeline Mode with Sparsity Support}
\label{subsection: sparsity support}
\noindent The pipeline mode for KAN operates in two stages to hide SPU latency, while performing sparsity-aware computation. In the first stage, each SPU fetches one input, while the SIMD core processes 16 inputs in parallel. To exploit the irregular sparsity introduced by the SPU, a zero-free data format~\cite{albericio2016cnvlutin} is employed. As shown in Fig.~\ref{Fig5:pipeline mode} (a), the TSE is divided into 16 slices, each storing non-zero outputs from an SPU in the dense data Scratchpad (Spad).

In the second stage, the Spads are grouped into two sets, providing dense input to multipliers in the PEs, and offsets are sent to the weight buffer to fetch corresponding weights. Therefore, the sparsity is utilized for skipping unncessary computation. Notably, the weight buffer dynamically adjusts its access scheme based on the operating mode. For KAN, two banks in the buffer are stacked as a group to deliver more weight data per batch for PE array, whereas in MLP mode, all four banks operate simultaneously to meet the bandwidth requirements for both the PE and SPU arrays (Fig.~\ref{Fig5:pipeline mode}(b)).

In addition to zero-free sparsity, user-configurable sparsity is supported. The TSE filters elements in batches of four, retaining only those specified by a mask defined during model training~\cite{fang2024maskllm, sun2023simple}. For example, with a pattern mask of 1 0 1 0, only nodes at indices 2'b00 and 2'b10 are retained if their values are non-zero (Fig.~\ref{Fig5:pipeline mode}(a)). This two-stage sparsity strategy can reduce the PE array’s computation by up to 87.5\%, enabling efficient accuracy scaling for KAN.

\vspace{-2mm}
\section{Experimental Results}
\label{section:experiment}

\begin{table*}[ht!]
\centering
\caption{Comparison of models on a real-world dataset}
\renewcommand{\arraystretch}{1.2} % 1.5倍行间距
\begin{tabular}{lcccccc}
\hline
Model & Layer Size & Para. Num. & Act. Config. & MSE (E-4) & RSE (E-1) & MAE (E-2) \\ \hline
MLP (4-layer) & {[}72,304,304,96{]} & 144k & ReLU (fixed) & 8.20 & 5.97 & 1.37 \\ 
MLP (3-layer) & {[}72,304,96{]} & 51k & ReLU (fixed) & 7.92 & 5.86 & 1.29 \\ 
KAN (3-layer) & {[}72,32,96{]} & 43k & silu \& B-spline: $G=4$ $K=3$ & \textbf{6.06} & \textbf{5.13} & \textbf{1.14} \\ 
KAN (2-layer) & {[}72,96{]} & 55k & silu \& B-spline: $G=4$ $K=3$ & 6.20 & 5.19 & 1.19 \\ \hline    
\end{tabular}
\label{table: model comparison}
\vspace{-2mm}
\end{table*}
In this section, we evaluate the VIKIN workflow in three steps: first, we create benchmark models based on a real-world dataset; second, we analyze performance gains from hardware optimizations, including the reconfigurable SPU and two-stage sparsity support; finally, we deploy the VIKIN prototype on an FPGA for overall evaluation.

\vspace{-2mm}
\subsection{Benchmark Model Setup}

\noindent To create benchmark models of KANs and MLPs for hardware evaluation, we conduct experiments on the Traffic dataset~\cite{lai2018modeling}, which contains road occupancy data collected in California during 2015–2016. This dataset is widely recognized as a benchmark for advanced time series forecasting models due to its complex temporal patterns and real-world relevance~\cite{cai2024msgnet, wu2021autoformer}. Following prior work~\cite{vaca2024kolmogorov}, we use traffic data from three consecutive days (72 hours) as input to predict traffic occupancy for the subsequent four days (96 hours). The dataset is split in a 7:2:1 ratio for training, validation, and testing, with all models trained for 100 epochs using the Adam optimizer and a learning rate of 0.001 to ensure a fair comparison.

Table~\ref{table: model comparison} summarizes the results of the trained models across various metrics, including Mean Squared Error (MSE), Root Relative Squared Error (RSE)~\cite{lai2018modeling}, and Mean Absolute Error (MAE). On this dataset, KAN demonstrates superior performance with the lowest error and only 29\% of the parameter count compared to the 4-layer MLP. This highlights the importance of a unified hardware framework capable of integrating both models for dynamic workloads on the edge.

\vspace{-2mm}
\subsection{Technique Evaluation}
\begin{figure}[h]
  \centering
  \includegraphics[width=0.95\columnwidth]{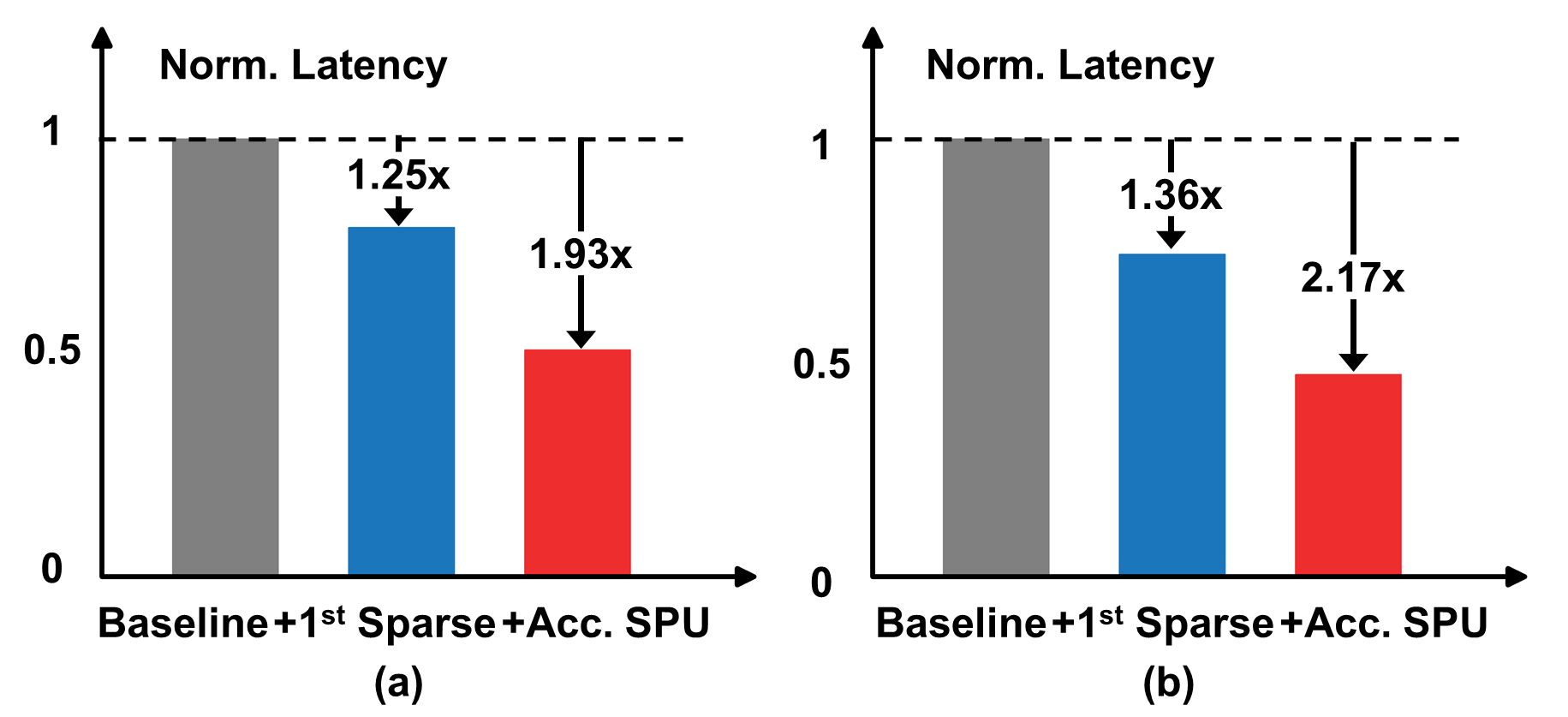}
  \caption{The performance contribution analysis of key techniques. (a) 3-layer MLP, (b) 4-layer MLP.
}
  \label{Fig:Reconfig_SPU performance gain}
\end{figure}

\noindent We deployed the trained models in Table~\ref{table: model comparison} on VIKIN to evaluate hardware performance. As shown in Fig.~\ref{Fig:Reconfig_SPU performance gain}, VIKIN accelerates the two MLP models by an average of 1.30$\times$ by simply skipping zero activations. When the SPU operates in accumulate mode to mimic the PE, the speedup increases to a maximum of 2.17$\times$ compared to the baseline, which is a simplified version of VIKIN without sparsity support and only the PE array is utilized.

\begin{figure}[h]
  \centering
  \includegraphics[width=0.90\columnwidth]{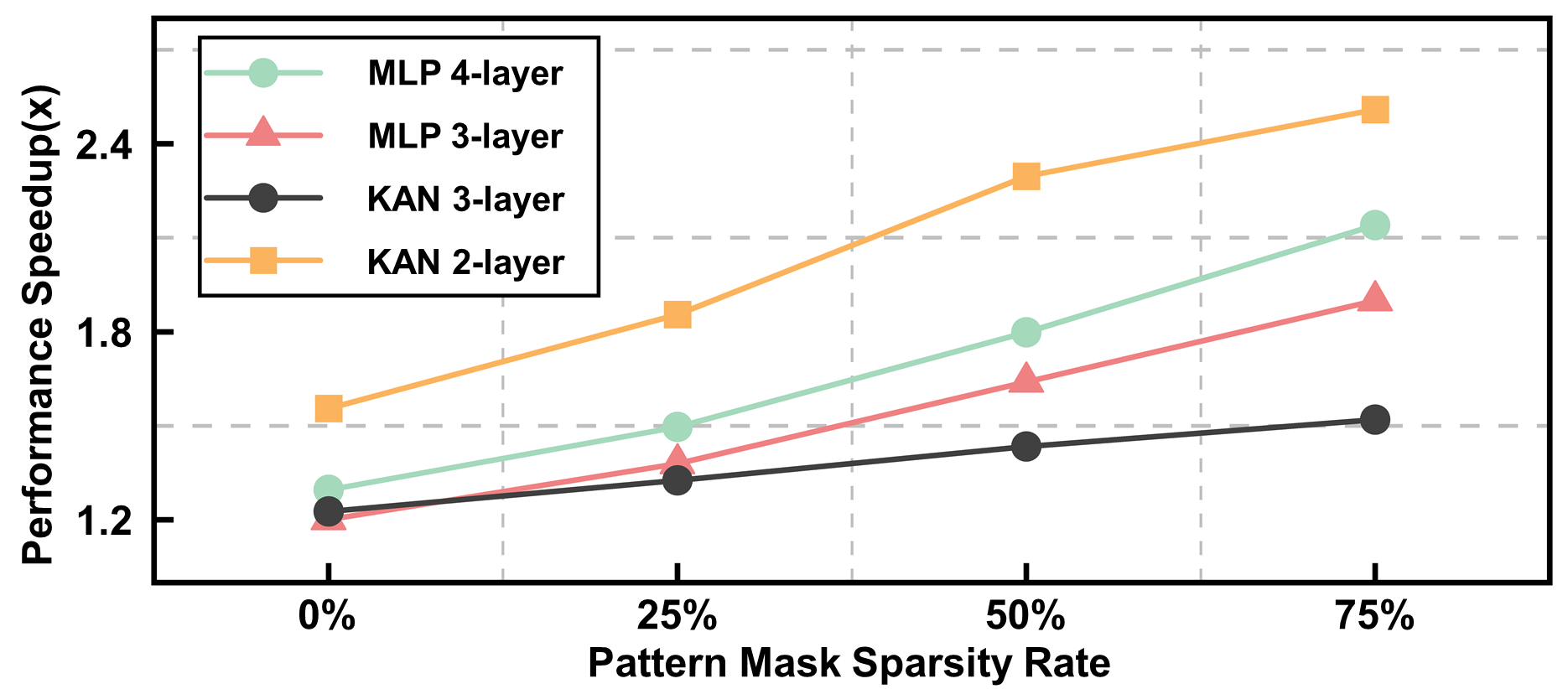}
  \caption{The speedup performance by two-stage sparsity}
  \label{Fig:double stage sparsity speedup}
\end{figure}

Fig.~\ref{Fig:double stage sparsity speedup} shows the performance gains from two-stage sparsity support over zero-free only, as discussed in Sec.~\ref{subsection: sparsity support}. With a pattern sparsity mask size of 4, the sparsity level can range from 0\% to 75\%. Even with the sparsity mask disabled (0\%), all models benefit from inherent sparsity. Enabling pattern sparsity furhter boosts performance, with the 2-layer KAN achieving up to a 2.50$\times$ speedup. KANs experience diminishing returns for higher sparsity rate due to a throughput mismatch between the PE and SPU arrays. Reducing values of $G$ and $K$ can improve this performance scaling.

\begin{figure}[h]
  \centering
  \includegraphics[width=0.95\columnwidth]{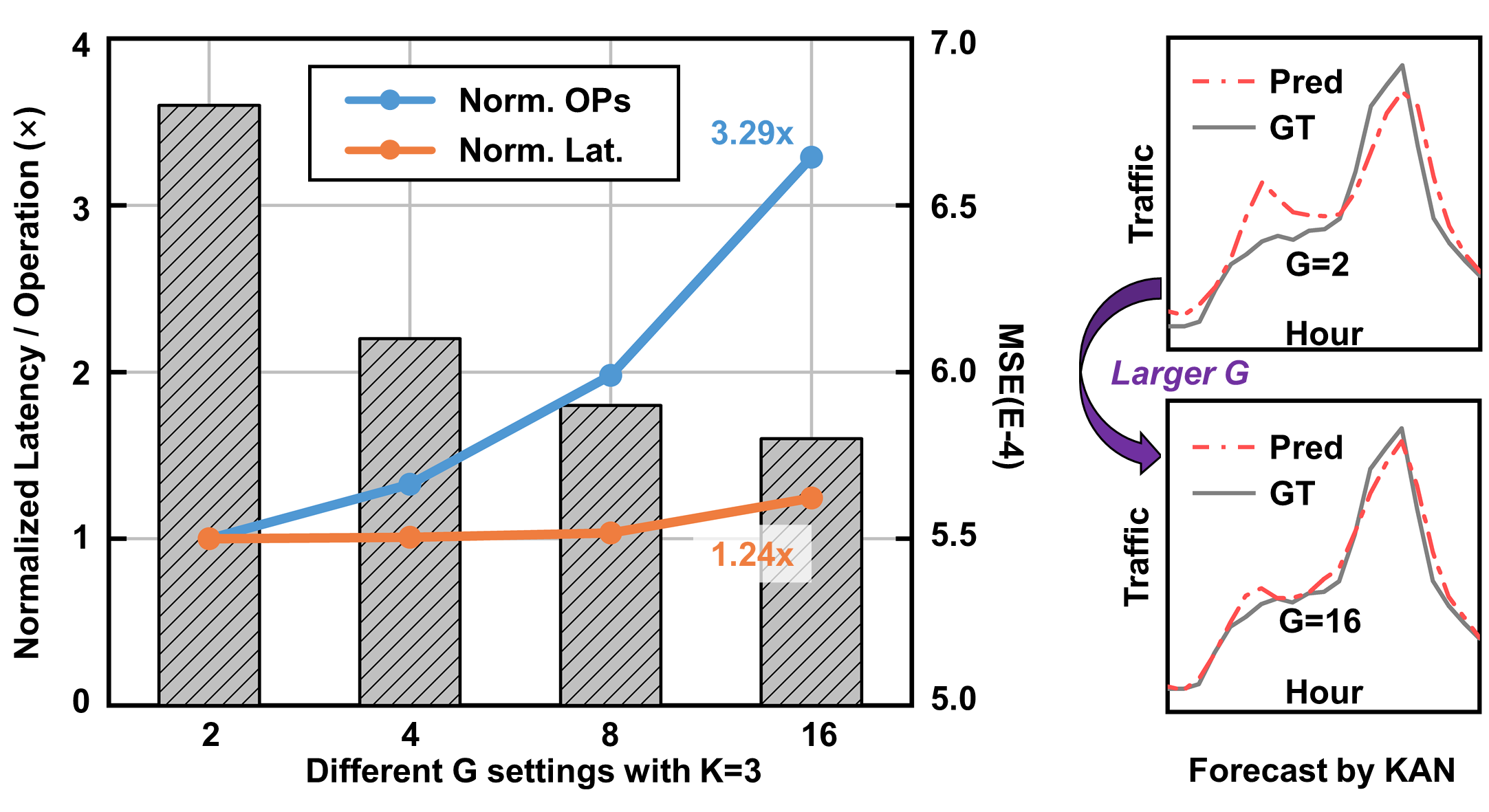}
  \caption{The performance analysis of KAN-specific parameters}
  \label{Fig:diff G}
\end{figure}

KANs allow for more grids (higher $G$) to improve accuracy, but at the cost of significant operation overhead. We further evaluated a 3-layer KAN with different $G$ values to analyze the impacts on hardware and algorithm performance (Fig.~\ref{Fig:diff G}). Higher $G$ improves detail fitting (lower MSE) but also incurs multiple times of operations. However, by leveraging the increased sparsity, VIKIN efficiently handles the more accurate model ($G=16$, 3.29$\times$ the operations of $G=2$) with only a 1.24$\times$ increase in latency.

\vspace{-2mm}

\vspace{-3mm}
\subsection{Overall Evaluation}
\noindent Table \ref{table: Overall Evaluation} shows the overall performance of VIKIN for executing two models, along with the hardware utilization of the Xilinx Virtex-7 VC709 operating at 115 MHz. Since the KAN model outperforms MLP in this application, VIKIN enables a 22\% reduction in latency by switching from MLP to KAN on the same hardware, leveraging KAN’s superior parameter utilization and efficient computation. Furthermore, the KAN model achieves significant gains with dedicated support on VIKIN, including a 1.25$\times$ speedup in throughput and a 4.87$\times$ improvement in energy efficiency compared to a common edge GPU with more advanced process technology.

\vspace{-4mm}
\begin{table}[htbp]
\centering
\caption{Overall evaluation of single-instance VIKIN.}
\renewcommand{\arraystretch}{1.2} % 1.5倍行间距
\renewcommand{\arraystretch}{1.2} % 调整行间距
\begin{tabular}{|l|c|c|}
\hline
\textbf{Model} & \textbf{KAN (2-layer)} & \textbf{MLP (3-layer)} \\ \hline
Dataset & \multicolumn{2}{c|}{Traffic \cite{lai2018modeling} (evaluate on FP16)}  \\ \hline
Patt. Spar. Rate & 50\% & 25\% \\ \hline
\begin{tabular}[c]{@{}l@{}}Error Metrics\& \\ Rel. Error \textcolor{gray}{$^{*1}$}\end{tabular}  
& \begin{tabular}[c]{@{}c@{}} MSE: 6.49 E$-4$ ($+5\%$) \\ RSE: 5.30 E$-1$ ($+2\%$) \end{tabular}
& \begin{tabular}[c]{@{}c@{}} MSE: 8.07 E$-4$ ($+2\%$) \\ RSE: 5.91 E$-1$ ($+1\%$) \end{tabular} \\ \hline
Latency (ms) & 7.47 E$-3$ & 9.56 E$-3$ \\ \hline
Speed-up \textcolor{gray}{$^{*2*3}$} & \textbf{1.25$\times$} & 0.72$\times$ \\ \hline
Ener. Effi. \textcolor{gray}{$^{*2*4}$} 
& \textbf{16.01 GOPS/W (4.87$\times$)} 
& 11.34 GOPS/W (2.20$\times$) \\ \hline
HW. Res. Util. & \multicolumn{2}{c|}{LUTs: 68529\:\:FFs: 83068\:\:DSPs: 340\:\:BRAMs: 82} \\ \hline
\end{tabular}
\begin{minipage}{0.49\textwidth} % 设置宽度为半栏
    \raggedright % 确保注释左对齐
    \vspace{0.3em}
$^{*1}$: The relative error is based on Table \ref{table: model comparison}.\\
$^{*2}$: The Jetson Xavier NX~\cite{nvidia2023jetsonxavier}, a common edge GPU with 21 TOPs peak performance, is used as the benchmark for both models in this task.\\
$^{*3}$: Speedup represents the performance improvement in average throughput.\\
$^{*4}$: Based on dynamic power measurements from the VC709 FPGA board.

\end{minipage}
\label{table: Overall Evaluation}
\end{table}

\vspace{-4mm}
\section{Conclusion}
\noindent This work presents VIKIN, a reconfigurable accelerator with two-stage sparsity support, for both KANs and MLPs efficient inference. VIKIN fully exploits the strengths of KANs to enhance real-time performance which is demonstrated on a practical real-world dataset. With reconfigurable pipeline and parallel modes, VIKIN can achieve significant performance gains by simply switching dataflow on the same hardware for case-by-case operations. 


\end{document}